\documentstyle[aps,prd,eqsecnum]{revtex}
\newcommand{\be}{\begin{equation}}
\newcommand{\ee}{\end{equation}}
\newcommand{\lll}{\langle}
\newcommand{\rrr}{\rangle}
\begin{document}

\title{ Higgs Mechanism
in Nontrivial Background. }
\author{V.I.Shevchenko\thanks{e-mail: shevchen@heron.itep.ru}}
\address{
{\it  Institute of Theoretical and Experimental Physics, \\
B. Cheremushkinskaya 25, 117218 Moscow, Russia}}
\maketitle
\begin{abstract}
The modification of classical dynamics
of abelian confining theory by virtue of
quantum Abrikosov-Nielsen-Olesen strings is discussed
taking D=4 abelian Higgs model as an example.
The form of string corrections to the
Wilson loop correlators, gauge boson
propagator, effective potential is presented
and possible relations to abelian-projected QCD
are outlined.
\end{abstract}

\section{Introduction}

 The confining properties of gauge theories attract considerable attention
for years. Despite quantum chromodynamics is undoubtedly
the most physically interesting example, analysis of other models
can also be instructive. Of particular interest
 is the scalar QED with the Higgs-type potential, the Abelian Higgs Model
(AHM) \cite{lg,abrikosov} -- the simplest abelian theory
exhibiting the property of confinement.
The dual charge and anticharge are confined in this
model already on the classical level through
the Abrikosov-Nielsen-Olesen (ANO) string formation.
\footnote{We consider the original and not the dual Abelian Higgs
model in
this paper, so the condensate is electrically charged
while magnetic monopoles are confined in the corresponding phase.}
It is numerically argued (see review \cite{obz}
and references therein),
that the dual version of this model might play
the role of effective theory of
abelian degrees of freedom for QCD after the
abelian projection procedure.
Vacuum is characterized by nonzero
expectation value of the scalar field
$\lll |\Phi|
\rrr$,
vector gauge boson is massive with the mass
$m_{\gamma}=e\lll |\Phi| \rrr$, providing Meissner effect,
 and Goldstone mode does not interact with the gauge
field. The Anderson--Higgs phenomenon is realized
in this way on the
classical level.

The quantum
spectrum of this theory includes excitations of
two different types.
The vector gauge boson, acquiring nonzero mass
$m$ due to
Higgs effect and
scalar particle with the mass $m_{\rho}$,
 corresponding to the excitation of the
condensate are of perturbative
type since their masses are proportional to the
corresponding couplings. The second class
of nonperturbative objects
is represented by the ANO--strings
and their excited states. Generally
speaking, one could find two
types of them: worldsheets of
classical (but, probably, vibrating)
strings bound by the external monopole loop (open strings)
and pure quantum excitations characterized by the
closed surfaces. Since the string tension of the ANO--strings
nonanalytically depends on the coupling, these excitations
have nonperturbative origin.

The consistent quantum theory of ANO--strings is unfortunately
still absent (see \cite{polyakov,polyakov1} in this respect). In particular,
it is not known, how to calculate the creation and annihilation
probabilities for the closed strings and also
the spectrum of quantum stringy states (which
are in some sense analogous to glueballs in QCD) cannot be determined.
Therefore exact dynamical calculations involving
ANO strings beyond perturbation theory are impossible.
Nevertheless one can extract some qualitative results
conserning the role quantum ANO strings can play in the dynamics of
the perturbative degrees of freedom. It is done in the
present paper by "phenomenological" parametrization
of the string contributions to different quantities
by one unknown function -- Gaussian string correlator.
In some cases this contribution is exact.


The partition
function we are going to consider reads \cite{lg,obz,orland,pol1,lee}:
\be
Z = \int {\cal D} A_{\mu} \; e^{-\frac14 \int F_{\mu\nu}^2\> d^4 x}\;
e^{-S[A]}
\label{eq1}
\ee
where
\be
e^{-S[A]} = \int {\cal D} \Phi \; e^{-\int\left( \frac12 |D_{\mu}[eA]\Phi|^2 +
\lambda (|\Phi|^2 - {\eta}^2)^2 \right) d^4 x}
\label{eq2}
\ee
We are working in 4d Euclidean space in the paper
and are not taking care of multiplicative field-independent factors
in all formulas.
The complex scalar field $\Phi(x) = \phi(x) e^{i\theta(x)}$ couples
with the gauge field $A_{\mu}(x)$ via covariant derivative
$D_{\mu}[eA]\Phi = {\partial}_{\mu} \Phi + ieA_{\mu}\Phi$.
The theory is assumed to be in the London limit
$\lambda / e^2 \gg 1$.
The radial part of the field $\phi(x)$ fluctuates
in the vicinity of $\eta$: $\Phi(x) = (\eta + \xi\rho(x))
e^{i\theta(x)}$, where $\xi \equiv 1/{\sqrt{\lambda}}$.
In the variables introduced eq.(\ref{eq2}) becomes:
\be
e^{-S[A]} = \int \rho {\cal D}{\rho}\> \int {\cal D}{\theta}\;
e^{-\frac{\xi^2}{2}\int ({\partial_{\mu}\rho})^2\> d^4 x
- \frac12 \int (\eta + \xi\rho(x))^2 (\partial_{\mu} \theta - eA_{\mu})^2\>
d^4 x
- \int (4\eta^2{\rho}^2 + 4 \xi\eta\rho^3
+ \xi^2 \rho^4)\> d^4 x}
\label{eq3}
\ee
The integration over $\rho$ gives at the leading order in $\xi$:
\be
e^{-S[A]} = \int {\cal D}{\theta}\;
e^{- \frac{m^2}{2} \int
(\partial_{\mu} \theta - eA_{\mu})^2\>
d^4 x
+ \frac{\xi^2 e^4}{16}\>\int
(\partial_{\mu} \theta - eA_{\mu})^4\>
 d^4 x}
\label{eq333}
\ee
We denoted classical mass of the vector gauge boson
$m=e\eta$ in eq.(\ref{eq333}).
As it is well known,
the $\theta$-dependence of the integrand
in (\ref{eq333}) cannot be omitted since for singular
configurations of the phase $\theta(x)$ it represents the
interaction between ANO-strings and the gauge field.
The Stokes theorem leads to the following relation
\cite{orland,pol1,lee}:
\be
{\partial}_{\mu} {\partial}_{\nu} \theta (x) -
{\partial}_{\nu} {\partial}_{\mu} \theta (x) = \pi
{\epsilon}_{\mu\nu\alpha\beta}{\Sigma}_{\alpha\beta}(x)
\label{eq12}
\ee
where ${\Sigma}_{\alpha\beta}(x)$ defines the
vorticity tensor current $
\Sigma_{\alpha\beta}(x) = \int d \sigma_{\alpha\beta}(\xi)
\delta^{(4)}(x(\xi)-x)
$
and integral is taken over some 2-complex $\Sigma$,
which physically is the world-sheet of the ANO string(s).
If there are no dynamical monopoles in the theory,
these world-sheets are closed: $ \partial_{\mu} {\Sigma}_{\mu\nu}(x)=0$.

\section{Correlators of the Wilson loops}

It is convenient to calculate Wilson loops
correlators for electric and magnetic
external currents, $j_{\mu}(x)$ and $J_{\mu}(x)$, respectively.
The electric current is chosen to be
$j_{\mu}(x) = \int\limits_C dz_{\mu}
\delta^{(4)}(z - x)$
while the magnetic one is introduced via the condition
$\partial_{\mu}\Sigma_{\mu\nu}(x) = J_{\nu}(x)$ where
$\Sigma_{\mu\nu}(x)$ is the vorticity tensor current
corresponding to the ANO-string
worldsheet.
It reflects the fact, that each monopole in the
condensate made of
electrically
charged particles must be accompanied by the ANO string.
Since $\partial\partial  \Sigma = 0$, correlators with total magnetic charge
not equal to zero are automatically excluded, as it should be
on the physical grounds.

The last term in the $S[A]$ in (\ref{eq333}), which is
proportional to the fourth power of the gauge fields does not
contribute at the leading order to the
bilocal current--current interaction.
It is suppressed by at least two powers of $\xi$ and by additional factors
coming from ANO-string correlation functions.

The Wilson loop for electric and magnetic currents
is given by:
\be
\lll W(C)\rrr =
\int {\cal D} A_{\mu} \; e^{-\frac14 \int F_{\mu\nu}^2\> d^4 x}\;
e^{-S[A]}\; e^{\> ie\zeta \int j_{\mu}A_{\mu} d^4 x  } =
\int {\cal D} \theta
\;e^{-\int d^4 x \int d^4 y\;
K(x,y)}
\label{eq18}
\ee
where
$$
K(x,y) =  {\pi}^2 {\eta}^2 {\Sigma}_{\mu\nu}(x)
{\Delta}_m(x-y) {\Sigma}_{\mu\nu}(y) +
 \frac{{\zeta}^2 e^2}{2} j_{\mu}(x)
{\Delta}_m(x-y) j_{\mu}(y) +
 \frac{1}{2}\left(\frac{2\pi}{e}\right)^2 J_{\mu}(x)
{\Delta}_m(x-y) J_{\mu}(y) +
$$
\be
+ i\pi \zeta j_{\mu}(x)\>
{\epsilon}_{\mu\nu\alpha\beta}\>
{\Sigma}_{\alpha\beta}(y)\>
\left[
\>\frac{\partial}{{\partial} x_{\nu}}\>\left(
\Delta_m(x-y) - {\Delta}_0(x-y)\right)\right]
\label{eq88}
\ee
where the massive propagator ${\Delta}_m(x-y)$ is given by
\be
{\Delta}_{m}(x) = \int \frac{d^4 p}{(2\pi)^4} \frac{e^{ipx}}{p^2
+m^2}
\label{eq6}
\ee
and $\Delta_0(x-y) = \Delta_{m=0}(x-y)$.

In case without external monopoles, i.e. $J_{\mu}(x)=0$,
eq.(\ref{eq88}) coincides with the result found in
\cite{pol1}. The correct factor for the charge
carried by the monopole currents manifests the Dirac
quantization condition. The factor $\zeta$ is the ratio
of the electric charge of the test particle
to the electric charge of the condensed particles,
which can be different from unity.

The integral over $\theta$ is taken
over the ensemble of all closed 2-complexes.
The nonlocality of the string action
in (\ref{eq88}) makes the correspondence
with known bosonic string
models, characterized by the local actions far from being
straightforward.

Nonperturbative effects in the model are taken into
account via the following string correlator
\be
\lll\lll {\Sigma}_{\mu\nu}(z)\>
{\Sigma}_{\rho\sigma}(t)\rrr\rrr
=
\frac{1}{Z}\;\int {\cal D} \theta
\;
 {\Sigma}_{\mu\nu}(z)\>
{\Sigma}_{\rho\sigma}(t)\;
e^{-  {\pi}^2 {\eta}^2 \int d^4 x \int d^4 y\;
{\Sigma}_{\mu\nu}(x)
{\Delta}_m(x-y) {\Sigma}_{\mu\nu}(y) }
\label{eq233}
\ee
with $\Sigma_{\mu\nu}(x)$ defined according to (\ref{eq12})
The Lorentz structure of the l.h.s. of (\ref{eq233})
is fixed in the unique way by the condition of
closeness of $\Sigma$:
\be
\lll\lll {\Sigma}_{\mu\nu}(z)\>
{\Sigma}_{\rho\sigma}(t)\rrr\rrr\> =
{\epsilon}_{\mu\nu\alpha\beta}
\frac{\partial}{{\partial} z_{\beta}}\>
{\epsilon}_{\rho\sigma\alpha\gamma}
\frac{\partial}{{\partial} t_{\gamma}}\;
d(z-t)
\label{eq244}
\ee
Such form of string correlator was explored in
\cite{zakh1}.

The expression (\ref{eq233}) is physically the sum over
all closed worldsheets of the ANO strings, which raises the question,
as it was already mentioned,
about possible calculation of it by string theory technique, using
$m_H^2=8\eta^2\lambda$ as a natural ultraviolet cutoff. It has not yet been
done in the continuum (see appendix of the paper \cite{pol1}
as an attempt in this direction).
 There are several points, which we would
like to discuss here. The first one is the following: it is easy to see,
that the group of transformations, leaving
eq.(\ref{eq233}) invariant is larger than
just reparametric one, and most of these symmetries are
hidden in some way.
In particular, any 2-complex in the
r.h.s. of eq.(\ref{eq18})
can be factorized into a linear superposition of 2d surfaces --
the property, which presumably does not hold in the nonabelian case.
It is not clear how one should
incorporate this symmetry into the measure ${\cal D} \theta$
if one wishes to rewrite it in the conventional
stringy form as integration over all embeddings and world-sheet
geometries \cite{polyakov}.

To construct the effective low-energy action for the abelian confining
strings
in the continuum
would be in its own turn
an interesting task. It is likely, that it would be local in the bulk, but not
on the world-sheet (see, for example, \cite{kleinert} where
Weyl--noninvariant string
action is proposed and crumpling is argued to be prevented by introducing
of higher order terms).

All this discussion is of academic interest, until it will help in
practical calculations with eq.(\ref{eq233}). The corresponding action is
nonlocal
and calculation of any nontrivial correlator starting directly from
eq.(\ref{eq233}) seems to be hopeless.
However, to guess how this correlator looks like, one can use field
theoretical
language. Introducing vector field $V_{\mu}$ via
the definition $\theta(x) = \int^x V_{\mu}dz_{\mu}$
and taking into account (\ref{eq12}) one gets
$$
\lll\lll {\Sigma}_{\mu\nu}(z)\>
{\Sigma}_{\rho\sigma}(t)\rrr\rrr
= - \frac{1}{Z}\;
\frac{\delta^2}{\delta J_{\mu\nu}(z)
\delta J_{\rho\sigma}(t)}\;
\int {\cal D} V_{\mu}\Gamma[V_{\mu}]
\; e^{-  \frac{{\eta}^2}{4} \int d^4 x \int d^4 y\>
{G}_{\mu\nu}(x)
{\Delta}_m(x-y) {G}_{\mu\nu}(y)} \cdot
$$
\be
\cdot e^{i\pi \int d^4 x  J_{\mu\nu}(x)
G_{\alpha\beta}(x)
{\epsilon}_{\mu\nu\alpha\beta}
}
\label{eq245}
\ee
where the field strength $G_{\mu\nu}(x) =
{\partial}_{\mu} V_{\nu}(x) -
{\partial}_{\nu} V_{\mu}(x)$
was introduced. The Jacobian $\Gamma[V_{\mu}]$ takes into account
compactness of the corresponding $U(1)$ and is proportional
on the lattice to the sum over all closed 1-cycles with some
weight:
\be
\Gamma[V_{\mu}] = \sum\limits_{\{ C \} } e^{\>i\oint\limits_{C}
V_{\mu}(z) dz_{\mu}}
\label{eq24}
\ee
Since these currents are formed by the electrically
charged quasiparticles,
which are condensed, we expect the sum in (\ref{eq24}) to be
divergent and
replaced in the continuum by
\be
\Gamma[V_{\mu}] = \int {\cal D} \chi \; e^{-\int\left( \frac12
|D_{\mu}[V]\chi|^2 + {\tilde L}(\chi) \right) d^4 x}
\label{eq255}
\ee
where the first factor in the exponent came from the naive
exponentiation of the corresponding determinant \cite{bardacki,halpern}
while the term  ${\tilde L}(\chi)$ provides the correct normalization
of the original action (\ref{eq2}).
Expression (\ref{eq255}) does not contain any small dimensionless
parameter.
If ${\tilde L}(\chi)$ is such, that the field $V_{\mu}$ acquires
mass, one gets in the infrared limit for the r.h.s. of (\ref{eq244})
\be
d(p)\sim \frac{1}{p^2 + M^2}
\label{eq266}
\ee
One expects to have such behaviour in the Higgs phase
of the theory, which was argued also in \cite{zakh1}, where
the mass $M$ was associated with the mass of the lowest-lying
$1^{-}$ "glueball" of this theory. It is easy to verify,
that in case $M=0$ the gauge
field $A_{\mu}$ becomes massless too -- this situation
corresponds to the
Coulomb phase.

\section{Gauge boson propagator and effective potential}

It is known, that the concept of propagator is of limited use
in the theories of the type considered here. The internal reason
is the presence of nonlocal interacting objects -- topological
defects, or, in other words, the fact, that Dirac
strings attached to the  monopoles become dynamical (i.e. just ANO strings,
see discussion in \cite{zakh2})
 due to the presence of the condensate.
The physical reason lies in impossibility to put into the
vacuum of such theory the probing test charge small
enough to neglect the corresponding
change in the properties of the
system. Born approximation does not work and the
potential nonanalytically depends on the charges of interacting
particles.

Therefore it is more instructive to consider gauge-invariant
correlation functions from the beginning, as we did.
Nevertheless
there is a possibility to interpret the results (at least partly)
in terms of gauge boson propagator.
It happens if the external currents
enter with a small parameter, the
role of which is played by $\zeta$ in (\ref{eq88}).

One should stress, that in the theory considered here (and
presumably in nonabelian theories too)
confinement is due to formation of the confining string(s) and
it is difficult (if not impossible) to describe this phenomenon
in terms of propagation of anything (see, for example, \cite{zakh1},
where the string formation was associated with the special
singularities of the propagator, and references therein).
On the other hand, one may consider only current-current terms
in the correlators like
calculated above
and extract the gauge boson propagator ${\cal D}_{\mu\nu}(x-y)$
from them
according to the following definition
\be
\lll W_{J \hat J}\rrr =
\lll \tilde W \rrr \cdot
e^{- \kappa \int d^4 x \int d^4 y\>
 J_{1}^{\mu}(x) {\cal D}_{\mu\nu}(x-y)
 {J}_{2}^{\nu}(y)}
\label{eq278}
\ee
where
the factor $\kappa$ is equal to ${\zeta}^2e^2$ or
$4\pi^2/e^2$ for $(jj)$
and $(JJ)$ cases, respectively and
the factor $\lll \tilde W \rrr$ accounts for the effects of
confinement and will not interest us below.
One gets from (\ref{eq88}) taking into account
(\ref{eq244}):
\be
{\cal D}_{\mu\nu}(p) =
\frac{\delta_{\mu\nu}}{p^2 + m^2(1-4\pi^2 {\eta}^2 d(p))} +
{\cal D}^{||}_{\mu\nu}(p)
\label{eq299}
\ee
It is worth noting, that one can use either
electric--electric or magnetic--magnetic correlators
 to derive (\ref{eq299}).
This could be expected from the beginning
 since despite there are two types of
particles (electrically and magnetically charged) in the theory,
there is only one photon.
The quantum correction proportional to $d(p)$
in (\ref{eq299}) leads to the deviation
of the static potential between electrical charges
from the classical Yukawa--type one; the classical potential
between external monopoles is also modified.
Exact lattice measurements of the static potentials
could shed some light on the relative importance
of the discussed effects, in particular, beyond the
London limit.

The stringy corrections  were discussed
in \cite{zakh1} and also in \cite{ant1}
for the propagator of the dual vector boson
(which mediates interaction between electric charges in
the magnetically charged monopole condensate),
while we are interested in the exchange between
electrically charged particles in the
condensate of electric charges.
To some extent the
dynamical corrections to the minimal string action
$$
S_{min} = \int d^4 p
{\pi}^2 {\eta}^2 {\tilde \Sigma}_{\mu\nu}(p)
{\Delta}_m(p) {\tilde \Sigma}_{\mu\nu}(-p) $$
are analogous to the quantities studied in 
\cite{zakh1}.\footnote{The dependence of the results
from \cite{zakh1} 
on the
auxilary vector $n_\mu$ inevitable in the theory with 
Higgs effect considered in the Zwanziger formalism
makes it difficult to 
compare (\ref{eq767}) to the results from
\cite{zakh1}.} 
The string fluctuations modify $\Delta_m(p)$ as
\be
\Delta_m(p) \to \frac{1}{p^2 + m^2 +4\pi^2 p^2 d(p)}
\label{eq767}
\ee
In the infrared region this correction
is damped as compared with (\ref{eq299}).
It is simple consequence of the fact, that the effective
vertex of the
Kalb--Ramond field (propagating according to (\ref{eq767}))
interaction contains additional power of momentum.
The signs of the corrections to (\ref{eq299}) and (\ref{eq767})
are different. For positive $d(p)$ it means, that
string fluctuations make the vector boson lighter while
the string tension smaller with respect the the classical
values.

In the strict sense the Gaussian approximation
adopted in (\ref{eq767}) is not justified contrary to
(\ref{eq299}), where it is controlled by the small
parameter $\zeta$.
There are special cases however where Gaussian correlator
provides exact answer. One example of such sort is the
effective potential.
Perturbative equation on extremum of the effective potential
reads (see, for example, \cite{malb})
\be
\frac{\lambda_R}{6} \rho^2 - m_R^2 = \frac{3e^4{\rho}^2}{16{\pi}^2}\>
\ln \> \frac{\mu^2}{e^2 {\rho^2}}
\label{iokko}
\ee
where $m_R$ and $\lambda_R$ are renormalized mass and coupling
constant
respectively  and $\mu$ -- normalization point.
The string contribution
has the following form:
\be
\frac{\delta S_{eff}}{\delta \rho} =
\frac{\lambda_R}{6} \rho^2 - m_R^2 - \frac{3e^4{\rho}^2}{16{\pi}^2}\>
\ln \> \frac{\mu^2}{e^2 {\rho^2}} + 12 {\pi^2} \int \frac{d^4 p}{(2\pi)^4}
\> \frac{p^4 d(p)}{(p^2 + e^2\rho^2)^2} = 0
\label{iou}
\ee
One expects, that the integral in the last term produces
quadratically divergent correction to the mass $m_R$, logarithmically
divergent contribution to $\lambda_R$ and finite term, having
the same structure as the r.h.s. of (\ref{iokko}).
 The resulting effective potential could have minimum therefore
at the point $\rho \neq 0 $  even if the strings are present.
The selfconsistent picture implies the string correlator $d(p)$
corresponding to the string solutions to the
equations following from the exact effective potential (\ref{iou}).

Having in mind the AHM as an effective theory for abelian--projected
QCD, one should mention, that there are numerical evidences
\cite{ilg}
that the parameters of the effective AHM correspond the Bogomolny
bound $m=m_{H}$ rather than in the London limit.
In particular, the QCD vacuum correlation length $T_g$ (which is
an analog of ${m}^{-1}$) is about 1 Gev while the lowest
excitation in gluodynamics is $0^{++}$ glueball, with the typical
mass 1.5 Gev. It means, that one {\it a priori} has no
reason to expect some kind of decoupling here since there are no degrees of
freedom,
which are really "heavy".

Therefore numerically observed facts,
supporting just {\it the classical} picture of confinement
in terms of AHM (for example,
exponential
profile of the chromoelectric field inside the QCD string)
should look rather unexpected if {\it quantum} corrections are
taken into account.
In other words, the classical picture is protected in the London
limit and not beyond it.
It should also be noted, that one cannot simply omit all quantum corrections
and consider classical equations of motion, following from AHM as the
only fingerprint of
the original nonabelian theory after abelian projection.
In particular, absence of integration over worldsheets in
(\ref{eq18}) for the case of single monopole loop $J$
would lead to the unphysical dependence of the
Wilson loop
on the arbitrary
chosen surface bound by $J$.

Another point is related with the fact, that
the string effective action has only one mass parameter
-- mass of the auxilary Kalb-Ramond field  (which coincides
with the vector boson mass).
The situation in QCD is much more complicated:
there is the tower of states, contributing to the
gauge--invariant 2-point
field strength correlator. It is an open question to what extent
the quantum dynamics of the abelian projected theory can
mimic this essentially  nonabelian feature.

\acknowledgements

The author is grateful to A.Dubin, M.Chernodub, F.Gubarev,
A.Morozov, M.Polikarpov and Yu.Simonov for useful discussions. The author also
wish to thank the {\it Dipartimento di Fisica} Univer\-sity of Pisa where the
part of this work was done and espe\-cial\-ly A.DiGiacomo for kind hospitality
and INFN for financial support.  The ICPM--INTAS grant 96-0457 is also
acknowledged.


\begin{references}
\bibitem{lg}
A.~Jaffe, C.~Taubes, {\it Vortices and monopoles}, Progr. in Physics, N2,
Birkhauser, 1980.
\bibitem{abrikosov}
A.A.~Abrikosov, Sov.Phys.-JETP {\bf 5} 1174 (1957);\\
H.B.~Nielsen, P.~Olesen, Nucl.Phys. {\bf B61} 45 (1973).
\bibitem{obz}
M.~Polikarpov, Nucl.Phys.Proc.Suppl.{\bf B53} 134 (1997);\\
M.~Chernodub, M.~Polikarpov, hep-lat/9710205.
\bibitem{polyakov}
A.~Polyakov, {\it Gauge fields and strings}, Harwood Academic
Publishers, 1987.
\bibitem{polyakov1}
A.~Polyakov, Nucl.Phys.Proc.Suppl. {\bf 68} 1 (1998).
\bibitem{orland}
P.~Orland, Nucl.Phys. {\bf B428} 221 (1994).
\bibitem{pol1}
E.~Akhmedov, M.~Chernodub, M.~Polikarpov, M.~Zubkov,
Phys.Rev. {\bf D53} 2087 (1996).
\bibitem{lee}
K.~Lee, Phys.Rev. {\bf D48} 2493 (1993).
\bibitem{zakh1}
M.Chernodub, M.~Polikarpov, V.~Zakharov, 
Phys.Lett.{\bf B457} (1999) 147.
\bibitem{ant1}
D.Antonov, I.J.Mod.Phys., {\bf A35} (1998) 214.
\bibitem{kleinert} 
M.~Diamantini, H.~Kleinert, C.~Trugenberger, hep-th/9810171.
\bibitem{zakh2}
F.~Gubarev, M.~Polikarpov, V.~Zakharov, hep-ph/9812030.
\bibitem{bardacki}
K.~Bardacki, S.Samuel, Phys.Rev. {\bf D18} 2849 (1978).
\bibitem{halpern}
M.B.~Halpern, P.Senjanovic, Phys.Rev. {\bf D15} 1655 (1977);\\
M.B.~Halpern, W.Siegel, Phys.Rev. {\bf D16} 2476 and 2486 (1977).
\bibitem{malb}
A.Malbouisson, F.Nogueira, N.Svaiter,
Mod.Phys.Lett., {\bf A11} (1996) 749.
\bibitem{ilg} 
F.Gubarev, M.Ilgenfritz, M.Polikarpov, T.Suzuki, Phys.Lett.
{\bf B468} (1999) 197.
\end{references}
\end{document}